# Recommendations for the Technical Infrastructure for Standardized International Rights Statements

*International Rights Statements Working Group, May 2015*

****Update 13 July**** **This paper is now closed for comments.** Thank you to everyone who has taken the time to leave comments. The Europeana & DPLA Working Group are reviewing the comments and will be incorporating them in the next version of the white papers. We will be providing further updates via our blogs.

# Introduction

This white paper is the product of a joint Digital Public Library of America (DPLA)-Europeana working group organized to develop minimum rights statement metadata standards for organizations that contribute to DPLA and Europeana. This white paper deals specifically with the technical infrastructure of a common namespace (rightsstatements.org) that hosts the rights statements to be used by (at minimum) the DPLA and Europeana.

These recommendations for a common technical infrastructure for rights statements outline a simple, flexible, and extensible framework to host the rights statements at rightsstatements.org. This white paper specifically outlines the management of rights statements as linked open data. The rights statements are published according to *Best Practices for Publishing RDF Vocabularies*.[1] They are encoded into dereferenceable URIs, express further information encoded in RDF, and link to existing vocabularies and standards. The rights statements adhere to expressions of existing rights vocabularies. Furthermore the paper reviews the publication and implementation to make the rights statements available through human-readable web pages augmented with machine-readable formats.

This document outlines the need for this common infrastructure, the Working Group's basic recommendations, and the reasons why we took this approach. This document is presented as a white paper, open for comments from the DPLA and Europeana communities. After a period of review of 45 days, the document and its recommendations will be finalized by the Working Group and published as a green paper.

[1] http://www.w3.org/TR/swbp-vocab-pub/

While the infrastructure proposed here is designed with DPLA and Europeana contributors in mind, we also recognize that the rights statements standards will be most useful if they are adopted even more broadly. They are therefore created with input from and thought to other cultural heritage organizations in both the North America and Europe. Comments from those organizations, and from organizations outside of the North America and Europe, are especially valued.

# Glossary of terms

- **Work:** a literary or artistic work of authorship contained in the collections of a cultural heritage institution.
- **Digital Object**: a digital representation of a work, or a born-digital work made available by a cultural heritage institution.
- **Public Domain**: Content, Metadata or other subject matter not protected by Intellectual Property Rights and/or subject to a waiver of Intellectual Property Rights.
- **License:** a legally binding license authorizing licensees to undertake specific actions that might otherwise be infringing on copyright and/or other intellectual property rights held by the licensor.
- **Rights Statement:** An assertion about the copyright status of a work drawn from a shared set of categories. In the rightsstatements.org DPLA-Europeana context, a rights statement is not a legal document per se, rather it is a simple, accurate, and unambiguous categorization that transparently describes a diverse set of legally-binding licenses.

# The Working Group

The DPLA-Europeana Rights Statements Working Group was created to develop a uniform set of copyright-status metadata elements for use primarily by cultural heritage institutions that contribute content and metadata to the Digital Public Library of America (DPLA) and to Europeana. The Rightsstatements.org governance structure, technical infrastructure, and metadata requirements outlined in this series of white papers has been produced by the Rights Statement Working Group in consultation with Creative Commons and with support from Kennisland.

**The Working Group consists of:**

Emily Gore – Working Group co-chair, DPLA, Director of Content
Paul Keller – Working Group co-chair, Kennisland, Coordinator Europeana Licensing Framework
Greg Cram – New York Public Library, Associate Director of Copyright & Intellectual Property
Julia Fallon – Europeana, IPR and Policy Advisor
Lucie Guibault – University of Amsterdam, Institute for Information Law
David Hansen – UNC School of Law, Assistant Clinical Professor & Reference Librarian
Antoine Isaac – Europeana, R&D Manager
Melissa Levine – University of Michigan Libraries, Lead Copyright Officer
Mark Matienzo – DPLA, Director of Technology
Patrick Peiffer – Bibliothèque Nationale de Luxembourg
Richard J. Urban – Florida State University College of Communication & Information, Assistant Professor
Maarten Zeinstra – Kennisland, advisor copyright and technology

**The Technical Working Group consists of:**
Valentine Charles - Europeana, Data R&D Coordinator
Esmé Cowles - University of California San Diego Library, Lead Developer
Karen Estlund - University of Oregon Libraries, Head, Digital Scholarship Center
Antoine Isaac – Europeana, R&D Manager
Tom Johnson – DPLA, Metadata & Platform Architect
Mark A. Matienzo – DPLA, Director of Technology
Patrick Peiffer – Bibliothèque Nationale de Luxembourg
Richard J. Urban – Florida State University College of Communication & Information, Assistant Professor
Maarten Zeinstra – Kennisland, advisor copyright and technology

# The Need for a Common Namespace

Europeana and the DPLA both have large collections of metadata pertaining to cultural objects. Libraries, archives, museums, and other cultural institutions have supplied each of these objects with a rights statement. In the case of Europeana, these statements are from a controlled list of rights statements.[2] However, the DPLA uses an open text field that has to date over 87,000 unique values.

[2] http://pro.europeana.eu/share-your-data/rights-statement-guidelines/available-rights-statements

The controlled list that Europeana maintains consists of all Creative Commons legal tools plus an additional set of Europeana made and controlled rights statements. These rights statements are dereferenceable URIs leading to human- and machine-readable representations of the rights statements following the "extended" requirements as described by *Best Practices for Publishing RDF Vocabularies*.[3]

Using a controlled set of rights statements makes metadata aggregated by projects such as Europeana and DPLA more valuable for reuse. It enables accurate search, and the use of algorithms to find works and communicate how works can be reused. It adds to the overall findability of the objects. It also allows explanations about the reuse of rights to be centralised. Working with a controlled set of statements also offers possible copyright options so that providers and copyright holders can provide more accurate assignment of appropriate statements for web resources of cultural objects. A controlled set of statements also provides a base for educational materials and tools to further educate the cultural heritage community about copyright.

As the Europeana rights statements are intended for the use in the Europeana dataset only, Europeana controls its language, versioning, and availability. This makes these statements less attractive to entities outside of Europeana, as they do not know if rights statements will remain the same or cause confusion about the range of applicability of the statements. In order for other parties to receive the same benefits of a controlled list of rights statements, we propose to develop a neutral namespace of rights statements: rightsstatements.org[4]. A neutral namespace for rights statements keeps the benefits of creating a controlled list, removes the drawbacks of a single controlling party, and enables the inclusion of additional contributors and partners in the future.

# URI Design

Rightsstatements.org will host persistent, dereferencable URIs that enable the delivery of human- and machine-readable representations of the rights statements.

The URI of the rights statement breaks down into the following components:[5]

- domain name (rightsstatements.org)

[3] http://www.w3.org/TR/swbp-vocab-pub/#requirements

[4] See also the white paper on the Governance of Rightsstatements.org and the white paper of the inital rights statements of Rightsstatements.org.

[5] Note that the presence of these components in the URI *should not* be considered as an expression of machine-readable metadata. Corresponding facts still need to be asserted at the data level. We consider URIs to be opaque for machine interpretation.

- resource type (rs)
- name of the statement (ic-donor-restrictions)
- version of the statement (1.0)
- jurisdiction where the statement applies using ISO 3166-1 (US) [optional]

A URI for the "In Copyright - Donor Restrictions (ic-donor-restrictions), US jurisdiction" rights statement would be:

```
http://rightsstatements.org/rs/ic-donor-restrictions/1.0/US/
```

# Data Modeling

A machine-readable description of the statements will be provided, which will be made available via the URI pattern above.

The group has made progress identifying vocabularies of classes and properties to express the data for the statements. Some basic principles have been agreed on:

- The working group will model the rights statement metadata using the Resource Description Framework (RDF) 1.1 Abstract Syntax[6] as a Simple Knowledge Organization System concept scheme.[7]
- The model will treat rights statements as members of the *dcterms:RightsStatement* class.
- The rights statements model requires all literals and/or lexical labels to include an appropriate language tag.

## Class for Rights Statements

In attempting to define classes for rights statements, the group identified an issue in current practice. Within the Europeana context, both Europeana Data Model and EU Rights Framework have adopted the Creative Commons Rights Expression Language (ccREL) *cc:License* class.[8] The CC REL RDF Schema asserts that a *cc:License* is a subclass of *dcterms:LicenseDocument (*"A legal document giving official permission to do something with a Resource"*)* which appears narrower than what is intended by the definition of *cc:License* ("a set of requests/permissions to users of a Work, e.g., a copyright license, the public domain, information for distributors").[9]

[6] http://www.w3.org/TR/2014/REC-rdf11-concepts-20140225/
[7] http://www.w3.org/TR/2009/REC-skos-reference-20090818/
[8] http://www.w3.org/Submission/ccREL/
[9] Further investigation reveals that is this partly due to early CC adoption of *cc:License* prior to the addition of classes to the DCMI Metadata terms in 2008.

Because rightsstatements.org rights statements are not legal documents per se, this group feels that using *cc:License* may be misleading, especially in cases that express public domain status. Therefore, this version of the rights statement concept scheme uses the broader *dcterms:RightsStatement* class ("A statement about the intellectual property rights (IPR) held in or over a Resource, a legal document giving official permission to do something with a resource, or a statement about access rights")[10].

## Relationships with Other Rights Statements

The set of rights statements provided by rightsstatements.org will be most valuable as Linked Data if it enables connections to other existing frameworks for expressing rights information. Whenever possible, a rightsstatement.org RDF representation will include references to related standards through the use of *skos:closeMatch, skos:exactMatch,* or *skos:relatedMatch.* For example, the PREMIS data model allows for the inclusion of a small set of coded rights status statements.[11] The rightssatement.org data can reflect this by including the following assertion:

```
rsdotorg:ic skos:relatedMatch premiscopy:cpr .
```

As the rightsstatements.org concept scheme develops, it will seek to incorporate relationships to other rights expressions and classifications deemed appropriate for the cultural heritage community.

## Open Data Modeling Issues

A number of modeling issues have still to be resolved by the group after feedback is received from the community.

### Property for Rights Statement Labels

Each rights statement will have a primary human-readable label ("In Copyright") and short identifier ("ic"). Related standards for expressing copyright status (CC, EDM) use *dc:title* while ODRS uses *rdfs:label*. In this version of the rightsstatements.org concept scheme, we propose using *skos:prefLabel* in line with our secondary goal of creating a SKOS vocabulary for the rights statements.

### Community Specific Permissions & Constraints

A feature of the proposed rights statements includes community specific permissions and constraints, for example "In Copyright - Educational Use Only." Using the Open Digital Rights Language (ODRL) Version 2.1 Ontology[12], we propose utilizing *odrl:permission* with *odrl:purpose* specifying "educational." As we could not identify an external vocabulary supporting “educational use” for this scheme, it would necessitate hosting and creating a term, e.g., http://rightsstatements.org/purpose/education. This larger question of hosting and maintaining community specific constraints in addition to rights statements at rightsstatements.org requires further discussion.

[10] http://purl.org/dc/terms/RightsStatement
[11] http://www.loc.gov/standards/premis/
[12] http://www.w3.org/ns/odrl/2/

#### Implementation & Extensibility

Because rights issues involve a number of additional facets beyond rights statements, the ability to provide guidance for implementation and determine whether the set of statements should be available as an extensible framework. Issues to be considered include:

- How to apply validity designations for a particular statement (e.g., “In Copyright” ending “2025-05-01” and “Public Domain” starting “2025-05-02”). Adding optional data to the URL might look like: http://rightsstatements.org/rs/pd/1.0/US/from/2025-05-02/ and http://rightsstatements.org/rs/ic-donor-restrictions/1.0/US/until/2025-05-01/
- Best practices on applying jurisdiction-specific restrictions to or with any statement that does not specify such restriction in its definition
- Best practices on applying additional rights and access related properties to objects, such as dcterms:rightsHolder and embargo restrictions

### Internationalization and Translations

Rights statements can be specific to a jurisdiction, but they are not language-specific. However we are in an international environment. The group may consider how to facilitate the creation of language-specific representations (translations) of a rights statement, with their own URI. Following the practice at Creative Commons, the URI for such a representation would have the URI elements mentioned above for the rights statement they represent, plus a filename or type component and a language component (".nl").

# Changes and Versions

Changes in human-readable text will require a new version of the rights statement. Minor editorial changes and changes in machine-readable elements will not necessitate a new version.

# Human and Machine readability

The rights statements vocabulary will contain a human- and machine-readable overview. Additionally, each rights statement will be available in human- and machine-readable versions. The human-readable version will be rendered in HTML generated by the RDF serializations. This section deals with the response a machine gets when a rights statement is requested. rightsstatements.org will offer the following formats:

- HTML5 with RDF(a)
- JSON-LD
- Turtle RDF syntax

The last two will be accessible through content negotiation using HTTP requests.

Machine readable formats are used to structurally communicate what the function of a web page is. Rights statements pages on rightsstatements.org will convey, at a minimum, the statement's title, descriptive and scope information, jurisdiction, creator, version, and other translations. Properties and classes will be drawn from the following namespaces:

| **Namespaces** | |
|---|---|
| Creative Commons Rights Expression Language (ccREL) | `http://creativecommons.org/ns#` |
| Dublin Core Elements 1.1 | `http://purl.org/dc/elements/1.1/` |
| DCMI Type Vocabulary | `http://purl.org/dc/dcmitype/` |
| DCMI Metadata Terms | `http://purl.org/dc/terms/` |
| Europeana Data Model | `http://www.europeana.eu/schemas/edm/` |
| ODRL | `http://www.w3c.org/community/odrl/two/vocab/2.1/` |
| PREMIS Copyright Status | `http://id.loc.gov/vocabulary/preservation/copyrightStatus/` |
| SKOS | `http://www.w3.org/2004/02/skos/core#` |

### Example in Turtle (RDF syntax)

The following example demonstrates the “In Copyright - Educational Use Only” rights statement expressed in RDF. This articulation is meant to illustrate how the rights statements might be expressed and is not a definitive version of the statement or properties applied.

```
@prefix dc: <http://purl.org/dc/elements/1.1/> .
@prefix dcmitype: <http://purl.org/dc/dcmitype/> .
@prefix dcterms: <http://purl.org/dc/terms/> .
@prefix edm: <http://www.europeana.eu/schemas/edm/> .
@prefix odrl: <http://www.w3c.org/ns/odrl/2/> .
@prefix premiscopy: <http://id.loc.gov/vocabulary/preservation/cop
@prefix rsdotorg: <http://rightsstatements.org/rs/> .
@prefix skos: <http://www.w3.org/2004/02/skos/core#> .

rsdotorg:ic-edu a dcterms:RightsStatement ;
    skos:prefLabel "In Copyright - Educational Use Only"@en ;
    skos:definition "This digital object is protected by
copyright and/or related rights. For educational uses, no
additional copyright permission is required from the copyright
holder. Please refer to the Data Provider for additional
information."@en ;
    dc:creator "Digital Public Library of America and Europeana
Rights Working Group" ;
    dcterms:hasVersion "1.0" ;
    dcterms:modified "2014-12-18" ;
    dc:identifier "ic-edu" ;
    skos:relatedMatch premiscopy:cpr ;
    odrl:permission [
       odrl:action odrl:use ;
       odrl:constraint [
          odrl:operator odrl:eq ;
          odrl:purpose
<http://rightsstatements.org/purpose/education>
       ]
    ] .
```

# Object-based Examples

This section outlines how an object may incorporate the rights statements from rightsstatements.org in its metadata.

### Object Available at Europeana

```
@prefix cc: <http://creativecommons.org/ns#> .
```

```
@prefix dc: <http://purl.org/dc/elements/1.1/> .
@prefix dcterms: <http://purl.org/dc/terms/> .
@prefix edm: <http://www.europeana.eu/schemas/edm/> .
@prefix ore: <http://www.openarchives.org/ore/terms/> .
@prefix rdfs: <http://www.w3.org/2000/01/rdf-schema#> .
@prefix skos: <http://www.w3.org/2004/02/skos/core#> .
@prefix foaf: <http://xmlns.com/foaf/0.1/> .

#  Example object at
http://www.europeana.eu/portal/record/92037/_http___www_bl_uk_online
gallery_onlineex_topdrawings_s_zoomify85637_html.html

# Equivalent to http://www.europeana.eu/rights/rr-f/
<http://data.europeana.eu/aggregation/provider/92037/_http___www_bl_
uk_onlinegallery_onlineex_topdrawings_s_zoomify85637_html> a
ore:Aggregation ;
     edm:aggregatedCHO
<http://data.europeana.eu/item/92037/_http___www_bl_uk_onlinegallery
_onlineex_topdrawings_s_zoomify85637_html> ;
     edm:dataProvider "The British Library" ;
     edm:isShownAt
<http://www.bl.uk/onlinegallery/onlineex/topdrawings/s/zoomify85637.
html> ;
     edm:provider "The European Library"@en ;
     edm:rights <http://rightsstatements.org/rs/ic/1.0/> .
<http://data.europeana.eu/item/92037/_http___www_bl_uk_onlinegallery
_onlineex_topdrawings_s_zoomify85637_html> a edm:ProvidedCHO ;
     dc:title "Stanton Harcourt, Church" ;
     dc:creator "Artist : Grimm, Samuel Hieronymus" ;
     dc:type "manuscript" ;
     dc:description "The 12th-century Church of St Michael contains
the tomb of Robert Harcourt, Henry VIII's standard bearer at the
Battle of Bosworth, 1485"@en ;
     dc:rights "Copyright © British Library Board"@en .
```

### Object Available at DPLA

```
@prefix cc: <http://creativecommons.org/ns#> .
@prefix dc: <http://purl.org/dc/elements/1.1/> .
@prefix dcterms: <http://purl.org/dc/terms/> .
@prefix dpla: <http://dp.la/about/map/> .
@prefix edm: <http://www.europeana.eu/schemas/edm/> .
@prefix ore: <http://www.openarchives.org/ore/terms/> .
```

```
@prefix rdfs: <http://www.w3.org/2000/01/rdf-schema#> .
@prefix skos: <http://www.w3.org/2004/02/skos/core#> .
@prefix foaf: <http://xmlns.com/foaf/0.1/> .

#  Example object at:
http://dp.la/item/fc69709e798f9ad881cf302953ad4c83

<http://dp.la/api/items/fc69709e798f9ad881cf302953ad4c83> a
ore:Aggregation ;
      edm:aggregatedCHO
<http://dp.la/api/items/fc69709e798f9ad881cf302953ad4c83#sourceResou
rce> ;
      edm:rights <http://rightsstatements.org/rs/ic-edu/1.0> .

<http://dp.la/api/items/fc69709e798f9ad881cf302953ad4c83#sourceResou
rce> a dpla:SourceResource ;
      dc:rights "Access to the Internet Archive's Collections is
granted for scholarship and research purposes only. Some of the
content available through the Archive may be governed by local,
national, and/or international laws and regulations, and your use of
such content is solely at your own risk" ;
      dc:creator "Boston Redevelopment Authority" ;
      dc:title "Educational institution study" .
```

### Object in Local Implementation[13]

```
@prefix dcterms: <http://purl.org/dc/terms/> .
@prefix lcnaf: <http://id.loc.gov/authorities/names/> .
@prefix premis: <http://www.loc.gov/premis/rdf/v1#> .
@prefix skos: <http://www.w3.org/2004/02/skos/core#> .
@prefix ucsd: <http://library.ucsd.edu/ontology/dams4.2#> .

<obj> dcterms:rights <http://rightsstatements.org/rs/ic-edu/1.0/>;
      premis:hasCopyrightJurisdiction "us";
      dcterms:accessRights ucsd:restrictedCampus;
      dcterms:rightsHolder lcnaf:n00085230 .
lcnaf:n00085230 skos:prefLabel "Doe, John, -1993" .
```

[13] Example from University of California San Diego Library

# Publication and Implementation

This section describes the proposed implementation for publishing the rights statements in both human- and machine-readable forms. Our recommendations follow the *Best Practice Recipes for Publishing RDF Vocabularies*[14], and address our requirements to provide access to these representations through content negotiation. Our choice of a specific recipe is informed by our need to satisfy all of the minimal and extended requirements as expressed in the *Best Practice Recipes*[15]*:*

- *M1. The 'authoritative' RDF description of a vocabulary, class, or property denoted by an HTTP URI can be obtained by dereferencing the URI of that vocabulary, class, or property.*
- *M2. The behavior of an HTTP URI denoting an RDFS/OWL vocabulary, class or property, does not lead to inconsistency in the interpretation of the nature of the denoted resource.*
- *E1. 'Human-readable' documentation about an RDF vocabulary, class or property, denoted by an HTTP URI, can be obtained by dereferencing the URI of that vocabulary, class or property.*
- *E2. Applications are able to differentiate between 'versions' of a vocabulary.*

In addition, we propose the introduction of two sub-requirements to requirement E1:

- *E1.1. A default translation of 'human-readable' documentation about an RDF vocabulary, class or property, denoted by an HTTP URI, can be obtained by dereferencing the URI of that vocabulary, class or property.*
- *E1.2. Additional translations of 'human-readable' documentation about an RDF vocabulary, class, or property, denoted by an HTTP URI, can be obtained by dereferencing the URI of that vocabulary, class, or property, along with the inclusion of an Accept-Language header in the request.*

As such, we propose the implementation based on a slight variation on recipe 5 from the *Best Practice Recipes* ("Extended configuration for a 'slash namespace', using multiple HTML documents").[16]

The proposed implementation will be an Apache HTTPD or Nginx web server, hosting static serializations of the rights vocabulary, and HTML-based representations for each individual translation. The HTML representations for each translation will be generated from the translations of literal values as expressed in the RDF vocabulary.

---

[14] http://www.w3.org/TR/swbp-vocab-pub/
[15] http://www.w3.org/TR/swbp-vocab-pub/#requirements
[16] http://www.w3.org/TR/swbp-vocab-pub/#recipe5

In addition to the web server configuration required by recipe 5, we will add configuration directives that support the use of the Accept-Language header. Additional requirement E1.1 will be satisfied by identifying the default translation, English, to return in the web server configuration directives. The following diagrams represent the additional behavior beyond recipe 5 as required by additional requirement E1.2:

Dereference the vocabulary URI, requesting HTML content in a specific language:

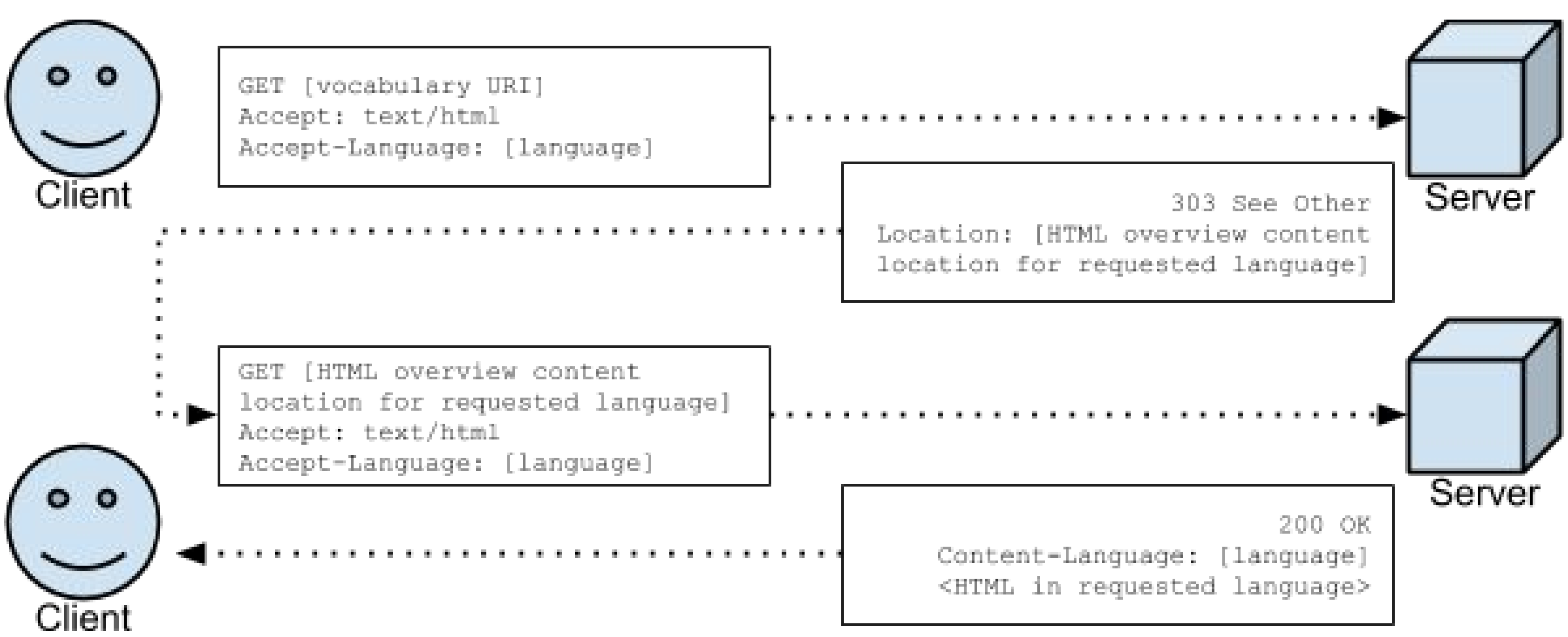

Dereference the URI of a class or property, requesting HTML content in a specific language:

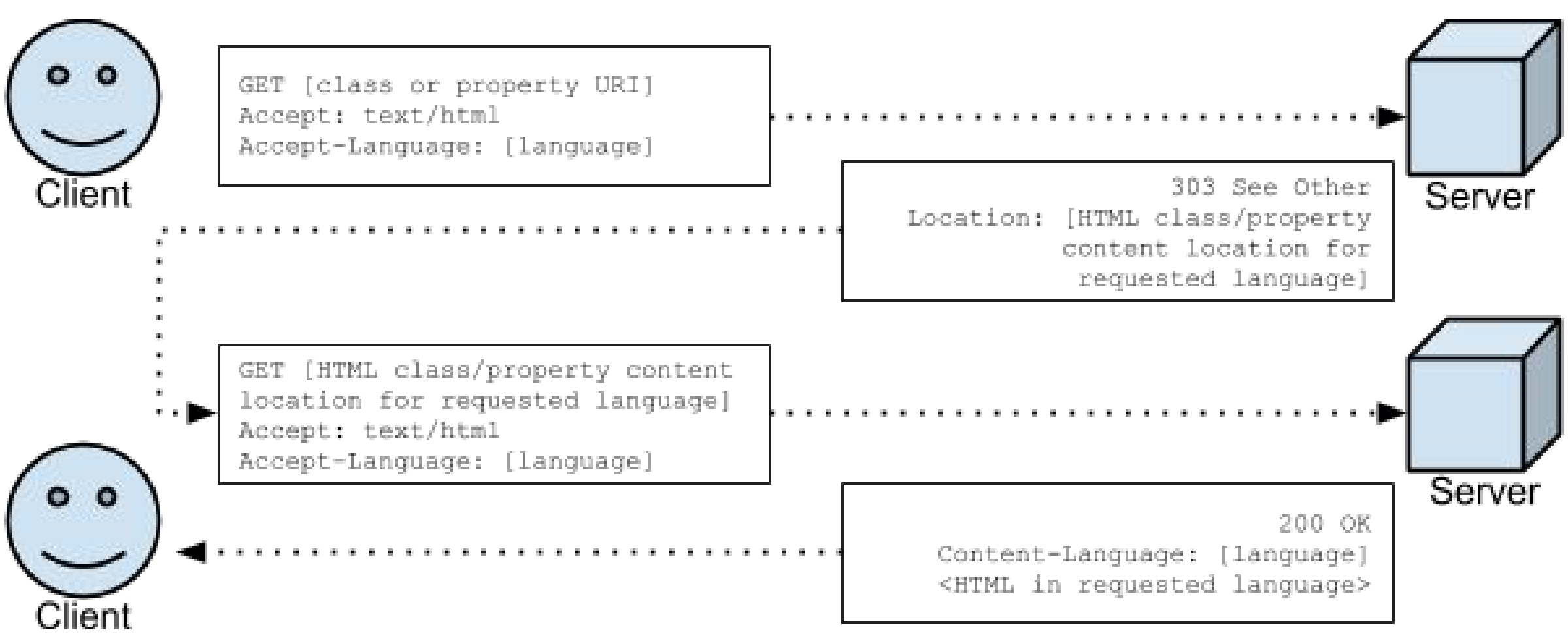

In addition to the web server configuration, we will develop scripts and documentation to support the transformation of the canonical RDF serialization (e.g., that are used for vocabulary maintenance) to the various RDF serializations and translations in HTML needed for publication.

In terms of management of the vocabulary, we feel that a minimal infrastructure involving a version control system such as Git (e.g., hosting on GitHub) will satisfy the immediate needs. While we anticipate the vocabulary changing, there remains an open question on governance about how and by whom new changes to the rights statements can be proposed. If more active, regular management is necessary, we propose the consideration of adopting a dedicated vocabulary management system.

# Acknowledgements

We are grateful to the following individuals who contributed during the public feedback period: Baxter Q. Andrews, Maarten Brinkerink, Nicholas Car, Leigh Dodds, Steven Folsom, Gloria Gonzalez, Kevin Hawkins, Jaffer, Lisette Kalshoven, Wibke Kolbmann, Sandra McIntyre, Aprille McKay, and Victor Rodriguez Doncel.